\documentstyle[12pt]{article}
\begin{document}
\newcommand{\newc}{\newcommand}
\newc{\be}{\begin{equation}}
\newc{\ee}{\end{equation}}
\newc{\ba}{\begin{eqnarray}}
\newc{\ea}{\end{eqnarray}}
\newc{\ie}{{\it i.e. }}
\newc{\eg}{{\it e.g. }}
\newc{\etc}{{\it e.t.c. }}
\newc{\etal}{{\it et al. }}

\newc{\ra}{\rightarrow}
\newc{\lra}{\leftrightarrow}
\newc{\no}{Nielsen-Olesen }
\newc{\lsim}{\buildrel{<}\over{\sim}}
\newc{\gsim}{\buildrel{>}\over{\sim}}

\begin{titlepage}
\begin{center}
             \hfill

{\large \bf
Power Spectrum of Cosmic String Perturbations
on the Microwave Background
}
\vskip 1cm
{\it Invited Talk at the conference `Unified Symmetry of the Small and the
Large', January 27-30 1994, Miami, Florida.}

\vskip .4in
{\large Leandros Perivolaropoulos}\footnote{E-mail address:
leandros@cfata3.harvard.edu}

{\em Division of Theoretical Astrophysics\\
Harvard-Smithsonian Center for Astrophysics\\
60 Garden St.\\
Cambridge, Mass. 02138, USA.}
\end{center}
\vskip .7in
\begin{abstract}
I review recent progress towards a detailed understanding of the power spectrum
of cosmic
string perturbations induced on the Microwave Background (CMB).
The use of a simple analytic model allows
to include important effects that have not been included in previous studies.
These effects are
potential fluctuations on the last scattering surface (the Sachs-Wolfe effect)
and Doppler CMB
perturbations. These previously neglected fluctuations are shown to dominate
over post-recombination
fluctuations (Gott-Kaiser-Stebbins effect) on angular scales less than about
$2^\circ$ assuming that
no reionization occurs. The effective power spectrum index $n_{eff}$ is
calculated on COBE scales and
is shown to be somewhat larger than 1 ($1.35 \lsim n_{eff} \lsim 1.5$). After
normalizing the model
on COBE data and string simulations, I derive the predicted values of
$({{\delta T}\over T})_{rms}$ for
several ongoing CMB experiments on medium and small angular scales.
A comparison is also made with current observations and predictions of models
based on inflation.
\end{abstract}
\end{titlepage}
\section{Introduction}
\par
One of the most important characteristics of a model for large
scale structure formation is the spectrum of primordial perturbations
the model predicts. This spectrum is imprinted on the last scattering surface
and parts of it show up on the cosmic microwave background (CMB) temperature
maps of experiments at various angular scales.

The CMB power spectrum predicted by models based on inflation\cite{gp82,bst83}
has been well
studied analytically\cite{bcd94} on all scales larger than 1-2 arcmin.
The main competition to these models comes from models where the primordial
perturbations
are generated by topological defects which produce seed-like primordial
perturbations. These
perturbations are imprinted on the CMB as temperature perturbations and can be
detected by ongoing
CMB experiments. A particularly interesting type of topological defect is the
{\it cosmic
string}.

The cosmic string theory \cite{k76,v85,b92} for structure formation is the
oldest and (together with
textures \cite{t89}) best studied theory of the topological defect class. By
fixing its single free
parameter $G\mu$ ($\mu$ is the {\it effective} mass per unit length of the
wiggly string and $G$ is
Newtons constant) to a value consistent\cite{tb86} with microphysical
requirements coming from GUT's,
the theory may automatically account for large scale filaments and sheets
\cite{v86,s87,pbs90,vv91,v92,hm93}, galaxy formation at epochs $z\sim 2-3$
\cite{bks87,as93} and galactic magnetic fields \cite{v92b}. It
can also provide large scale peculiar velocities \cite{v92a,pv93} and is
consistent with the
amplitude, spectral index\cite{bt86,bbs88,vs90,bsb92,p93a,p94} and the
statistics\cite{g90,p93b,mpb93,cfg93}
 of the cosmic microwave background (CMB) anisotropies measured by the COBE
collaboration
\cite{s92,w92a} on large angular scales ($\theta\sim
10^\circ$).

The CMB
 spectrum of the cosmic string model has been investigated using both
simulations\cite{bbs88,bsb92} and analytical methods\cite{p93a} but only on
large angular scales (\ie scales of a couple of degrees or larger).
 This paper shows the results of an attempt I have recently made\cite{p94} to
extend
the analysis of the CMB spectrum predicted by cosmic strings to arbitrarily
small scales. Due to resolution limitations this extension can only be made
using
analytical methods.

In the next section I will briefly review the
 three basic mechanisms by which strings can induce perturbations
on the CMB: post-recombination Gott-Kaiser-Stebbins\cite{ks84,g85} (GKS) type
perturbations,
potential  perturbations on the last scattering surface (LSS) and Doppler
perturbations
due to
velocities of electron last scatterers\cite{e89}. Then
I will describe an analytical method\cite{v92a,p93a,p93b,p93c,pv93}
used to derive the power spectrum of CMB fluctuations induced by strings, in
terms of the single free parameter of the model ($G\mu$) and three evolution
parameters that can be fixed from string simulations. In section 3, I will show
how can these
parameters be fixed by using the COBE detection for the free parameter $G\mu$
and string simulations for the evolution parameters.
The scale invariance of the spectrum will also be demonstrated. Finally, in
section 4,
I will use the
resulting normalized spectrum to make predictions for the rms temperature
fluctuations detected by ongoing CMB experiments on medium and small angular
scales.

I will assume $\Omega_0 = 1$ $h=1/2$, Cold Dark Matter (CDM), $\Lambda=0$ and
standard recombination.

\section{Types of CMB Fluctuations}

The best studied mechanism for producing temperature fluctuations on the
CMB by cosmic strings is the Gott-Kaiser-Stebbins (GKS) effect\cite{ks84,g85}.
According to this
effect, moving long strings present between the time of recombination $t_{rec}$
and today produce (due to their deficit angle\cite{v81})
 discontinuities in the CMB temperature between
photons  reaching the observer through opposite sides of the string.

The second mechanism for producing CMB fluctuations by cosmic strings is based
on potential fluctuations on the LSS. Long strings and loops present between
the
time of equal matter and radiation $t_{eq}$ and the time of recombination
$t_{rec}$ induce density
and  velocity fluctuations to their surrounding matter. These fluctuations grow
gravitationally and at $t_{rec}$ they produce potential fluctuations on the
LSS.
Temperature fluctuations arise because photons have to climb out of a potential
with spatially dependent depth.

The third mechanism for the production of temperature fluctuations is based on
the
Doppler effect. Moving long strings present on the LSS produce velocity fields
to
the surrounding plasma. Thus, photons scattered for last time on these
perturbed
last scatterers suffer temperature fluctuations due to the Doppler effect.

The total temperature perturbation may be obtained by superposing the effects
of these three mechanisms at all times from $t_{rec}$ to today. Clearly each
mechanism involves the superposition of some type of seed present during a
given period of time. Therefore, in order to construct the spectrum of the
resulting perturbations we must address the following two questions: First,
{\it how can we superpose the seeds in order to construct the resulting
spectrum}
and second {\it what is the type of seed corresponding to each mechanism for
producing CMB fluctuations?}

I will first address the first question. Consider a great circle on the sky
(\eg a
meridian) and a seed function $f_1^\Psi (\theta)$ of angular scale $\Psi$,
superposed
$N$ times at random  positions $\theta_n$ on the circle with variable
magnitudes
$a_n$. The resulting pattern will be
\be
f(\theta)=\sum_{n=1}^{N} a_n f_1^\Psi
(\theta-\theta_n)={1\over{2\pi}}\sum_{n=1}^{N} a_n
\sum_{k= -\infty}^{+\infty}{\tilde f_1^\Psi}(k) e^{ik(\theta-\theta_n)}
\ee
where ${\tilde f_1^\Psi} (k)$ is the Fourier transform of $f_1^\Psi (\theta)$
\begin{equation}
{\tilde f_1^\Psi}(k)\equiv \int_{-\pi}^{+\pi} d\theta f_1^\Psi (\theta) e^{-i k
\theta}
\end{equation}
 Therefore ${\tilde f} (k)$ (\ie the Fourier transform of $f (\theta)$) can
be expressed in terms of ${\tilde f_1^\Psi} (k)$
\be
{\tilde f}(k)={\tilde f_1^\Psi}(k)\sum_{n=1}^{N}a_n e^{i k \theta_n}
\ee
and the corresponding power spectrum $P_0 (k)$ is easily obtained as the
ensemble
average of $|{\tilde f}(k)|^2$.
\be
P_0 (k)\equiv <\vert {\tilde f}(k)\vert^2>=N \vert{\tilde f_1^\Psi}(k)\vert^2
<a_n^2>
\ee
 The $1\sigma$ error to $P_0 (k)$ can also be
obtained from the standard deviation of $a_n^2$. In realistic cases the
superposed seed functions $f_1^\Psi (\theta)$ will be perturbations induced by
topological defects which obey a {\it scaling solution} and therefore their
size
is a fixed fraction of the horizon at any given time. This implies that the
size
of the superposed seeds will be larger at later times due to the expansion of
the
comoving horizon. Thus when the comoving horizon grows by a factor $\alpha$,
the
size of the superposed seeds will grow by the same factor while their total
number $N$ will be reduced by $\alpha$ since the total number of horizons
included in the circle will be smaller at that time. By considering $Q$
expansion
steps for the comoving horizon we obtain
\be
P_Q (k)=\sum_{q=0}^Q P_q (k)\equiv \sum_{q=0}^{Q} {N\over {\alpha^q}} \vert
{\tilde f_1}^{\alpha^q
\Psi} (k) \vert^2 <a_n^2>
\ee
where $Q$ is determined by the maximum and minimum size of the superposed seeds
(or the corresponding horizon scales) as
\be
Q={{\log ({{\Psi_{max}}\over {\Psi_{min}}})}\over {\log \alpha}}
\ee
while $N$ is the product of the number of seeds per horizon $M$ times the
number
of horizon scales included in the circle during the first expansion step when
the
horizon scale is $\Theta(t_i)$
\be
N(t_i)=M\times {{2\pi}\over \Theta(t_i)}
\ee

In order to obtain the power spectrum in the case of strings we must define the
types of seed functions $f_1(\theta)$ that need to be superposed and the
initial
and final times of superposition for each mechanism generating perturbations.

A long string moving with velocity $v_s$ between the LSS and an observer will
induce, according to the GKS effect a temperature discontinuity of
magnitude\cite{ks84,g85,s88}
\be
{{\delta T}\over T}= 4 \pi G\mu (v_s \gamma_s)_{rms} {\hat k}\cdot ({\hat v}
\times {\hat s})
\ee
where ${\hat k}\cdot ({\hat v}\times {\hat s})$
is a geometric factor which depends on the relative
orientation between the string ${\hat s}$ and the photon unit wavevector ${\hat
k}$.
 Therefore an observer scanning the sky along
an {\it arc} that itersects the string will detect a temperature step function
with amplitude ${{\delta T}\over T}$ and angular scale $2\Psi$ depending on the
string curvature radius $\xi$ given as a fraction of the angular size $\Theta$
of
the horizon at the time the photons interacted with the string ($\Psi(t)\equiv
\xi \Theta(t)/2$).

Moving long strings present between $t_{eq}$ and $t_{rec}$ produce velocity
perturbations towards the surface they sweep in space. Assuming CDM, these
perturbations grow rapidly and form planar density enhancements called {\it
wakes}. Let $\sigma(t_i,t_{rec})$ be the surface density of a wake present at
$t_{rec}$ and formed at an earlier time $t_i$. The wake, due to its surface
density will induce potential fluctuations at a distance $h$ from its surface.
The potential fluctuations in turn produce temperature fluctuations on the
photons departing from the LSS, with magnitude\
\be
{{\delta T}\over T}={1\over 3} \Phi_w(x,t_{rec})={1\over 3}4\pi
G\sigma(t_i,t_{rec}) |h(x)|
\ee
and size $\Psi$ determined by the scale of coherence of the long string that
produced the wake. In (9) $h(x)=x\hspace{1mm} \cos\phi$ is the {\it
perpendicular}
distance to the wake as a function of
the angular distance $x$ from the wake on the last scattering surface.  The
{\it angular} distance from the wake over which this perturbation persists is
approximatelly $\Psi(t_{i})$.Therefore, an observer scanning the sky along an
{\it arc}
that intersects a wake will detect a temperature fluctuation pattern described
by equation (9)
as a function of the angular scale $x$.

String simulations have shown that the string network contains in addition to
long strings, a component of tiny loops with typical sizes about $10^{-4}$ the
size of the horizon at any given time. A loop present on the LSS will also
induce
potential perturbations due both to its energy density and to the dark matter
it
has accreted. The corresponding temperature fluctuation may be approximated by
a
function that is 0 on scales much larger than the size of the loop and equal to
${{\beta G\mu}\over 3} ({t\over t_i})^{2/3}$ on smaller scales \ie
\begin{eqnarray}
{{\delta T}\over T}&=&{1 \over 3}\Phi_l (x,t)\simeq {{\beta G\mu}\over 3}
({t\over t_i})^{2/3}
\hspace{1cm} \vert x\vert \leq R \\
{{\delta T}\over T}&\simeq & 0 \hspace{1cm} \vert x\vert > R
\end{eqnarray}
The time dependent factor represents the gravitational growth of the loop
induced
perturbation on the CDM and $\beta$ is a parameter of order smaller than 10
determining the length of the loop as a function of its radius $R$. Since the
typical angular size of a loop present at $t_{rec}$ is 1 arcsec or less, we
expect the
effects of loops to be negligible for all present experiments since their
resolution is on much larger scales. This is verified from the results shown
below.

The velocity perturbations induced by long strings present on the LSS produce
velocity fields on the electrons last scatterers of CMB photons. These fields
induce temperature fluctuations by the Doppler effect. The important
perturbations are those produced on the plasma {\it at} $t_{rec}$ since earlier
perturbations are damped by photon drag and pressure effects.
The magnitude of the produced temperature fluctuations is equal to the
projection
of the plasma velocity with respect to the observer on the photon unit
wavevector
\be
{{\delta T}\over T}={\hat k}\cdot {\vec v}=\lambda \pi G\mu v_s \gamma_s {\hat
k}
\cdot ({\hat v}\times {\hat s})
\ee
where $\lambda$ is defined as
\be
\lambda=(1+{{(1-{T\over \mu})}\over {2 (v_s \gamma_s)^2}})
\ee
and is different from 1 in the case of a wiggly string in which the tension $T$
is not equal to the mass per unit length $\mu$. It measures the Newtonian
interaction with matter induced by the wiggles of the string. Simulations
indicate $\lambda \simeq 6$. The scale of these Doppler perturbations is
$\Psi$,
determined by the coherence length of long strings present on the LSS.

 Having specified the types of seeds that need to be superposed, it is now
straightforward to use (5) to obtain the power spectrum component induced by
each
type of seed. The total spectrum is obtained by summing up the spectrum
components.
\be
P(k) = P_{GKS} (k) + P_W (k) + P_L (k) + P_D (k)
\ee

A typical spectrum component obtained for the GKS term is
\be
P_{GKS}(k)={{(4\hspace{1mm}\pi \hspace{1mm}G\hspace{1mm}\mu)^2 \hspace{1mm}
<(v_s \hspace{1mm}\gamma_s)^2)> \hspace{1mm} 200\hspace{1mm}
M}\over 3}
 \sum_{q=0}^{Q=23}
{{16\hspace{1mm}\sin^4(\xi \hspace{1mm}0.008 \hspace{1mm}
\hspace{1mm}\alpha^q\hspace{1mm} k)}\over
{\alpha^q k^2}}
\ee
The part $(4\hspace{1mm}\pi \hspace{1mm}G\hspace{1mm}\mu)^2\hspace{1mm}
<(v_s \hspace{1mm}\gamma_s)^2)>$ comes from the amplitude of the GKS step
function (8), the $1/3$ comes by averaging over all string orientations, the
$200M$ is the number of
seeds included in the first expansion step occuring at $t_{rec}$ and the sum is
over the Fourier transform of the seed function scaled to expand with the
comoving horizon at each expansion step.
The assumption that each expansion step occurs when the physical horizon
doubles in size implies
that $\alpha \simeq 2^{1/3} \simeq 1.26$. Decreasing $\alpha$ has the effect of
increasing the
number of terms $Q$ in the sum (15) but decreasing the value of each term. Thus
we expect that (15)
should not be very sensitive to the precise value of $\alpha$.

 The other spectrum components (Wakes,
Loops and Doppler) have similar forms expressed in terms of sums\cite{p94}.
They depend on four parameters: the only free parameter $G\mu$, the parameter
$b\equiv M \hspace{1mm} <(v_s \gamma_s)^2>$, the string coherence length
(curvature radius) $\xi$ as a fraction of the horizon scale
$\Theta (t)$ ($\Psi (t) = \xi \Theta (t)/2$ and the parameter
$\lambda$ determined by the wiggliness of the string. The three evolution
parameters $b$, $\xi$ and $\lambda$ may be fixed by comparing with numerical
simulations while the free parameter $G\mu$ is fixed by comparing with
observations (\eg the COBE detection).

\section{\bf Fixing Parameters}

In order to make predictions about ongoing CMB experiments we must determine
the
only free parameter $G\mu$ as well as the parameters $b$, $\lambda$ and $\xi$.
String simulations \cite{bb88,as90} indicate that
$M\simeq 10$ while $(v_s\gamma_s)_{rms} \simeq 0.15$ implying $b\simeq 0.24$
and $\lambda\simeq 6$.
I will verify these values by directly fitting our spectrum with partial CMB
spectra obtained by
simulations on large angular scales. Bouchet, Bennett and Stebbins\cite{bbs88}
(hereafter BBS) have
used numerical simulations to calculate the term $P_{GKS} (k)$ for a single
expansion step. Their
result for the total power on angular scales smaller than $\theta_*$ with
$t_i=t_{rec}$, $t_f\simeq 2
\hspace{1mm} t_{rec}$ is
\begin{eqnarray} P_{BBS} (\theta \leq
\theta_*,\Theta_i=\Theta_{rec})&=&\int_{2\pi/\theta_*}^\infty
{{d^2k}\over{(2\pi)^2}} P_{BBS} (k)
\nonumber\\  &=&(6G\mu)^2
({{\theta_*^{1.7}}\over{0.0012+\theta_*^{1.7}}})^{0.7}
\end{eqnarray}
\newpage
The present analysis, focusing on a line across the sky rather than a patch
predicts
\begin{equation}
P_{an}(\theta \leq \theta_*,\Theta_i=\Theta_{rec})=2\hspace{1mm}
\int_{2\pi/\theta_*}^\infty
{{dk}\over{(2\pi)}} P_{KS}^{Q=0}(k)
\end{equation}

\vspace{10cm}
{\large \bf Figure 1:}The total power on scale less than $\theta_*$ produced by
cosmic strings
during one expansion step starting at $t_{rec}$.
\vskip .5cm

Figure 1 shows $P_{an}(\theta_*)$ for $b=0.237$ and $\xi=0.45$ (continous line)
superimposed on
$P_{BBS}(\theta_*)$ (dashed line). It also shows the $1\sigma$ errors to
$P_{an}(\theta_*)$ obtained
from the variance of $a_n^2$. The values $b=0.237$ and $\xi=0.45$ were chosen
in order to obtain
the best fit to $P_{BBS}$ but they are in very good agreement with the expected
values
(obtained for $M\simeq 10$, $(v_s \gamma_s)_{rms}\simeq 0.15$ and string radius
of curvature about
half the horizon scale).

Figure 2a shows a superposition of the components of the spectrum (\eg equation
(15) for the GKS
term) with the above choice of parameters. The sums were performed using {\it
Mathematica}\cite{w91}.
Clearly the GKS term (continous line) dominates on large angular scales
($\theta >
4^\circ$) while the Doppler term (long dashed line) is dominant on smaller
scales. The contribution of
potential perturbations by wakes (dotted line) is less important but is clearly
not negligible
especially on scales of a few arcmin ($k\simeq 1500$). Finally, the contibution
of loops (short dashed
line) is negligible on all scales larger than 2-3 arcmin ($k\leq 8000$). Figure
2b shows the product
$kP(k)$ fot the total spectrum (equation (14)) with $1\sigma$ errors denoted by
the dotted lines.

\vspace{10cm}
{\large \bf Figure 2a:} The four components of the power spectrum of CMB
perturbations induced by
cosmic strings
\vskip .5cm

One of the most interesting questions that may be addressed using the spectrum
of Figure 2b is
{\it `What is the effective power spectrum index n, predicted by cosmic strings
on COBE angular
scales?'}. Previous studies\cite{bsb92,p93a} have
addressed this question without taking into account the effects of potential
and Doppler
perturbations. The correlation function $C_1(\theta)$ for perturbations along a
great circle is given
in terms of $P(k)$ as
\begin{equation}
C_1 (\theta)=<{{\delta T}\over T}(\phi) {{\delta T}\over
T}(\theta+\phi)>_\phi={1\over{(2\pi)^2}}\sum_{k=-\infty}^{k=+\infty} P(k) e^{i
k \theta}
\end{equation}

\vspace{10cm}
{\large \bf Figure 2b:} The {\it total} power spectrum of string perturbations
along a great
circle on the sky.
 \vskip .5cm

For 2d maps the corresponding equation is ($l\gg1$, $\theta \ll \pi$)
\cite{e89}
\begin{equation}
C_2(\theta)\simeq{1\over{(2\pi)^2}}\int d^2l \hspace{1mm} C_l \hspace{0.5mm}
e^{i {\vec l}\cdot {\vec \theta}}
\end{equation}
By isotropy we must have $C_1 (\theta)=C_2 (\theta)$. It may also be shown that
$l^2 C_l\sim
l^{n-1}$ where $n$ is the power spectrum index. Since both $k$ and $l$ are
Fourier conjugate of
$\theta$ we have $k\simeq l$.
\newpage
Also (30) and (31) imply (with $\theta\simeq 0$) that $P(k)\simeq \pi l C_l$
and
\begin{equation}
kP(k)\sim k^{n-1}
\end{equation}

\vspace{10cm}
{\large \bf Figure 3a:} The best linear fit to the total power spectrum for $5
\leq k \leq 20$.
\vskip .5cm

Figures 3a and 3b show the best linear fit of the log-log plot $kP(k)$ vs $k$,
for $5\leq k\leq 20$
and $5\leq k\leq 100$ respectively.
The best fits give $n=1.35$ (Figure 3a) and $n=1.48$ (Figure 3b). This result
indicates that cosmic
strings favor values of $n$ somewhat larger than 1 in agreement with recent
indications from the
Tenerife experiment and the second year data of COBE \cite{b94} which favor
$n\simeq 1.5$. In contast it is much harder for inflationary models to explain
such high values of
$n$ \cite{ll94,s93}.

There is a simple analytic way to show that in the sector of the power spectrum
where the
GKS effect dominates, a scale invariant ($n\simeq 1$) spectrum should be
expected.
For ${\tilde f_1^\Psi}(k)\sim \Psi {\tilde f_1^{\Psi=1}}(k\Psi)$ (as in the
case of the
GKS seed functions), (12) may be writen as
\begin{equation}
P(k)=\sum_{q=0}^Q P_q (k)=\sum_{q=0}^Q \alpha^q P_0 (\alpha^q k)\simeq \alpha
P(\alpha k),
\hspace{0.3cm} \Psi_{max}^{-1}\leq k \leq \Psi_{min}^{-1}
\end{equation}

\vspace{10cm}
{\large \bf Figure 3b:} The best linear fit to the total power spectrum for $5
\leq k \leq 100$.
\vskip .5cm

Therefore
\begin{equation}
k P(k)\sim {\rm const}
\end{equation}
which indicates a scale invariant spectrum for the Kaiser-Stebbins term in the
angular scale range
$\theta \geq 2^\circ$ ($k\leq \Psi_{min}^{-1}$). The Kaiser-Stebbins term
plotted in Figure 2a
(continous line) is in agreement with this result (the best fit for this
component of the spectrum
is obtained for $n=1.12$).

\section{Predictions}
 The predicted ${{\delta T}\over T}_{rms}$ for ongoing experiments can only be
obtained after the
free parameter $G\mu$ is fixed. This may be achieved by using the COBE
detection. Using the DMR
filter function $W(k)\simeq
e^{-k^2/18^2}$ and comparing the observed ${{\delta T}\over T}_{rms}=1.1 \pm
0.5$
with the predicted one
\be
{{\delta T} \over T}_{rms}=(C(0))^{1/2}=[{1\over {2\pi^2}}\sum_{k=0}^\infty
P(k) W(k)]^{1/2}
\ee
the single free parameter $G\mu$ is fixed to the value
\be
G\mu\simeq 1.6\pm 0.5
\ee

\vspace{10cm}
{\large \bf Figure 4:} The cosmic string predicted correlation function
smoothed on COBE scales. Superimposed are the first year COBE data.
\vskip .5cm

Having completely fixed the spectrum we are now in position to use appropriate
filter functions to
make predictions for ongoing CMB experiments\cite{expr} on various scales.
Figure 4 shows the correlation function obtained by Fourier transforming the
power spectrum
and smoothed by the COBE filter function. Superimposed are the COBE
data\cite{s92,w92a}.

Table 1 shows the
approximate location of the maxima ($k_0$) and the spreads ($\Delta k$) of the
filter functions of
several ongoing experiments. It also shows the coresponding detections or upper
bounds. The cosmic
string predictions were obtained by using (23) with gaussian filter functions
of the form
\be
W(k)=e^{-(k-k_0)^2/{\Delta k}^2}
\ee
and the power spectrum shown in Figure 3.
The predictions of the standard model based on inflation ($0.8\leq n \leq 1.0$,
$\Lambda=0$, CDM)
have been derived in Ref. \cite{bcd94}. From Table 1, it is clear that the
predictions of both the
standard CDM model and cosmic strings are consistent with detections at the
$1\sigma$ level.
However this may well change in the near future as the quality of observations
improves.

{\bf Table 1}:Detections of ${{\Delta T} \over T}_{rms}\times 10^{6}$ and the
corresponding
predictions of the string and inflationary models ($0.8\leq n \leq 1.0$,
$\Lambda=0$) normalized on
COBE. \vskip 0.1cm
\begin{tabular}{|c|c|c|c|c|c|}\hline
Experiment & $k_0$ & $\Delta k$ & Detection & Strings & Inflation \\ \hline
COBE & 0 & 18 &11 $\pm$ 2&11 $\pm$ 3  & 11 $\pm$ 2 \\ \hline
TEN& 20 & 16 &  $\leq$  17 &13 $\pm$ 3  & 9 $\pm$ 1\\ \hline
SP91& 80 & 70 & 11 $\pm$ 5 &20 $\pm$ 5  & 12 $\pm$ 2 \\ \hline
SK&  85 & 60 & 14 $\pm$ 5 &19$\pm$ 4   & 12 $\pm$ 3 \\ \hline
MAX& 180 & 130 & $\leq$  30 ($\mu Peg$) &21 $\pm$ 5.5   &16 $\pm$ 5   \\ \hline
MAX& 180 & 130 & 49 $\pm$ 8 ($GUM$) &21 $\pm$ 5.5  & 16 $\pm$ 5  \\ \hline
MSAM &  300 & 200 & 16 $\pm$ 4   &19 $\pm$ 4    & 24 $\pm$ 6   \\ \hline
OVRO22 &  600 & 350 & - &13 $\pm$ 4    & 17$\pm$ 7  \\ \hline
WD&  550 & 400 &  $\leq$  12  & 17.5 $\pm$ 4.5    & 7 $\pm$2   \\ \hline
OVRO&  2000 & 1400 &  $\leq$  24 & 13.5 $\pm$ 3.5   & 7 $\pm$ 3  \\ \hline
\end{tabular}
\vskip 0.5cm

In conclusion, I have demonstrated, using a simple analytical method, that the
CMB spectrum
predicted by the cosmic string model can be calculated in a straightforward way
including all the
relevant sources of perturbations. I have also shown that the results are
consistent with numerical
simulations even though their validity extends beyond the resolution of present
simulations.
Finally I showed that the predicted power spectrum index is slightly larger
than 1
($n_{eff}\simeq 1.4$) and that the predicted {rms} temperature fluctuations
${{\Delta T} \over
T}_{rms}$ are consistent with detections to this date on all angular scales
larger than 2-3
arminutes.

This analysis has assumed standard recombination and values of cosmological
parameters
($\Omega_0=1$, $h=1/2$, CDM, $\Lambda=0$). It is important to extend these
results to less standard
cases including reionization or presence of Hot Dark Matter. Work in this
direction is in progress.
\vspace{0.5cm}

{\bf Acknowledgements}

\noindent
 This work was supported by a CfA Postdoctoral Fellowship.


\end{document}